\documentclass[conference]{IEEEtran}
\IEEEoverridecommandlockouts
\usepackage{cite}
\usepackage{amsmath,amssymb,amsfonts}
\usepackage{algorithmic}
\usepackage[linesnumbered,ruled,vlined]{algorithm2e}
\usepackage{graphicx}
\usepackage{textcomp}
\usepackage{xcolor}
\usepackage{multicol}
\usepackage{multirow}
\usepackage{hyperref}
\usepackage{tabularx}
\usepackage{makecell}
\usepackage{subfig}

\usepackage{lineno,hyperref}
\usepackage{adjustbox}
\usepackage{threeparttable}

\def\BibTeX{{\rm B\kern-.05em{\sc i\kern-.025em b}\kern-.08em
    T\kern-.1667em\lower.7ex\hbox{E}\kern-.125emX}}
\begin{document}

\title{Using Game Theory to maximize the chance of victory in two-player sports\\}

\author{\IEEEauthorblockN{Ambareesh Ravi\IEEEauthorrefmark{1}, 
Atharva Gokhale\IEEEauthorrefmark{2}, 
Anchit Nagwekar\IEEEauthorrefmark{3}}
\IEEEauthorblockA{
\textit{Department of Electrical and Computer Engineering} \\
\textit{University of Waterloo, Waterloo, CANADA}\\
Email: \IEEEauthorrefmark{1}ambareesh.ravi@uwaterloo.ca,
\IEEEauthorrefmark{2}asgokhal@uwaterloo.ca,
\IEEEauthorrefmark{3}ahnagwek@uwaterloo.ca}}

\maketitle

\begin{abstract}
Game Theory concepts have been successfully applied in a wide variety of domains over the past decade. Sports and games are one of the popular areas of game theory application owing to its merits and benefits in solving complex scenarios. With recent advancements in technology, the technical and analytical assistance available to players before the match, during game-play and after the match in the form of post-match analysis for any kind of sport has improved to a great extent. In this paper, we propose three novel approaches towards the development of a tool that can assist the players by providing detailed analysis of optimal decisions so that the player is well prepared with the most appropriate strategy which would produce a favourable result for a given opponent's strategy. We also describe how the system changes when we consider real-time game-play wherein the history of the opponent's strategies in the current rally is also taken into consideration while suggesting.
\end{abstract}

\section{Introduction}
There have been various advancements in the way in which technology is utilized in sports and games. The aid provided to players through various technological means has improved at a rapid pace over the past few years. Match analysis results in a high volume of statistical data which in turn is used by players and coaching team for match preparation. In this paper, we propose to use a game-theoretic approach to develop a tool that can potentially help players in \textit{any two-player sports} to thoroughly investigate the opponent and prepare a strategic plan to maximize the chance of winning.

For the purpose of proposing a viable and feasible solution that can be directly employed in the real-world, we have chosen \textit{badminton} as the primary sport for experiments, analysis and discussion owing to our deep understanding, the common passion for the sport and since it is easy to understand. Badminton is a racket sport in which two players alternately hit a shuttlecock until a point is scored by one of them. In this work, we propose two models and both the models use the history of match data of the opponent and the player under consideration as the input to provide suggestions and recommendations for a player. We also discuss the necessary steps to develop a complete, end-to-end solution integrating different types of data and technology in order to create a dedicated software application/ program that can be customized for each sport.

Our first model called \textit{the recommendation system} takes into consideration the various shots that the player and the opponent have played throughout their career. The main purpose of this system is to help the players gain an understanding of the different shots that the opponent plays and to gain knowledge of the best possible shots that he/she can play such that the chances of gaining a point are maximum. This model could be used by the players and coaches before going into a match and is not intended to be used during match play. Our second model which we like to call the \textit{Simulation model} is an extension of the recommendation system but it takes into consideration the history of the match when it is in use. We use a reward system to determine the best shots for the players. This model is intended to provide match practice to the players against opponents so that they can simulate match situation and gain experience before heading into the actual match.

Also, we intend to utilize some of the recent revisions in Badminton laws, one of which allows for coach intervention during the match. The proposed tool can prove to be very handy here as the coach can influence the player and guide the player in the middle of the match through the use of this tool by quickly analysing the match up till that particular point in the game-play. Moreover, our method could prove valuable in saving players and coaching staff the huge amount time expended in going through hours of match videos of the player and opponents and can critical help them in quantitatively analyse the performance and recommend the necessary preparation from players historical data.

\section{Relevant works}
Game theory has been used to study various strategic sports in the past and most of the prevalent work done so far has been towards studying specific parts of a sport and not the entire match or game. In soccer, the penalty kick has been modelled as a strategic game with imperfect information because of uncertainty about the kicker's type \cite{penalty}. Bayesian equilibrium concept was used and it was found out that the kickers adopt a mixed strategy equilibrium depending based on their strong foot. In cricket, a normal form game was modelled between the batsman and the bowler \cite{Cricket}. The strategies of the batsmen depended upon the type of shot played and the bowler's strategies were the different types of deliveries that he can bowl. The utility values were derived based on the probability of the player to take a particular strategy. The study revealed that the probability distribution followed by the players in adopting different strategies in real-world cricket is very close to the Nash equilibrium conditions. Alpha Go, an artificial intelligence entity that can play the game of Go which was developed by Deep-Mind was able to defeat the best player in the world by 5 games to 0. This intelligent program uses a combination of Monte Carlo Simulation with value and policy networks \cite{Alpha} with the concept of Markov Decision Processes (MDP) \cite{Burnetas:1997:OAP:265654.265664}\cite{Alpha}\cite{vanOtterlo2012} as its base. 

The best of the players in any sports around the world learn at a rapid pace as they progress in their professional careers. However, there are few areas where this natural learning process doesn't prove to be very effective. In such cases, advanced mathematical and computer modelling can come to aid and convert this slow time-consuming process into a rapid and results-oriented one. In our literature survey, we came across many instances where game theory was used to solve problems pertaining to sports. One such case traces back to 2012 in the Olympics encounter between Yu Yang and Wang Xiaoli of China and South Korean pairs Jung Kyun-eun/Kim Ha-na and Ha Jung-eun/Kim Min-jung (doubles). The Chinese team tried to lose on purpose in this group stage encounter to avoid playing against their teammates  Tian Qing and Zhao Yunle so that China is assured both the gold as well as the silver medals. \cite{HONGZHI20131222} presents a detailed analysis on badminton match throwing using this example through game theory. The study reveals that the reason for this kind of match throwing lies in the loopholes of the format that the competition adopts and any rational player would adopt this strategy in the interest of the team. Besides this, there have been various other similar cases that drove us towards using game theory for our problem. In \cite{Pollard}, game theory is used to determine the optimal time during the match to play a risky serve and how the surprise factor plays a part, also studying how it affects the outcome for the player under consideration. Apart from that, it is found that it is beneficial for the player to play a risky serve during the critical points of the match rather than the less important ones. Highly motivated with the past work along these lines, through this project, we intend to contribute towards the game of badminton and develop a highly effective tool for player assistance with the aid of game theory concepts. 

In summary our contributions are:
\begin{enumerate}
    \item A recommendation tool to suggest the best shots for each of the possible shots of the opponent using the concept of \textit{best strategy} from game theory.
    \item A simulation model that considers the history up to two shots while determining the favorable shot to be played at any particular stage of the match and the approach is modelled as \textit{an approximated finite non-zero sum extensive from game}
\end{enumerate}

\section{Data Collection}
Comprehensive data is essential to effectively model the capability and choice of players which is vital in sports. Badminton is not like board games where predefined moves or strategies can consistently help you win. Humans tend to think differently when it comes to physical capabilities especially in sports. We can’t expect a player to play the same shot with the same accuracy every time; rather, we understand that a player’s capability, stability and mentality changes during the course of a game. But to best model these factors, data plays a key role.

Our data was manually collected by going through several full match videos of the players. We considered matches between two of the best and most consistently performing badminton players in the world – Lin Dan from China and Lee Chong Wei from Malaysia so that the inefficiency of the players won't tamper with the final results and also incorporate the variation of left-handed and right-handed player in our data. The matches we recorded are spanned over a period of 8 years (2011 - 2019) so that we cover the changing game plan and shot selection over a considerable period. The data was manually collected and annotated on a shot-by-shot basis for a comprehensive modelling, with their outcomes in terms of points and sets won. This format is essential to calculate the necessary parameters for the proposed models. 

\begin{figure*}
    \centering
    \includegraphics[width=0.75\textwidth]{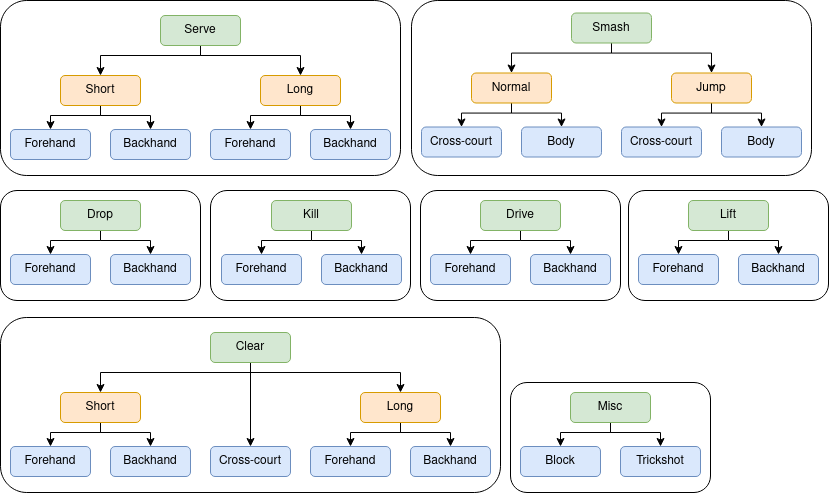}
    \caption{Various badminton Shots under consideration}
    \label{Badminton shots}
\end{figure*}

The types of shots which we have considered while collecting our data contribute to efficient results. Figure \ref{Badminton shots} presents the scope of the shots for our paper. We have tried to incorporate the position of the player on the court in terms of the type of shots. Also, there are certain exceptions to the types of shots that can be played against a particular shot of the opponent. None of the shots can be responded with a service which is quite obvious. Also, a smash and a block cannot be returned by a smash and a block respectively. Drops are usually quite difficult to smash. It is important to note that important factors like agility, fatigue and mental state of the player during the course of the game are not taken into account while modelling due to the complexity involved.

\section{Methodology}

The detailed workings of each of the proposed methods are discussed in this section along with the mathematical modelling. For the purpose of modelling and easy representation, the player under consideration who uses our proposed approaches is referred to as $X_p$ and the player's opponent as $X_o$

\subsection{Recommendation Tool}
The recommendation tool considers $X_p$’s choice and capabilities based on $X_p$'s history with a particular opponent $X_o$ to offer the best suggestions to each shot of $X_o$. The concept of best response from game theory is adopted which will help $X_p$ to be match ready with the best and safe shots to play given any shot from $X_o$ to maximize $X_p$'s chances of winning each point and in turn winning the match. We model the recommendation tool as a \textit{normal form game}. The reason being that we are only worried about $X_p$; meaning, we only care about $X_p$'s strategy and how to maximize $X_p$'s chances of winning the match and not $X_o$'s. So it is enough to consider the game on a shot-by-shot basis rather than a sequential game. At any stage of the game, for a given shot $s_{-i}$ of $X_o$, $X_p$ will have set of probable shots (strategies) $S$ to play and the recommendation tool outputs those best shot $s_{i}*$ from the available shots $S$ which is the best response to $s_{-i}$. The game can be modelled as a tuple representing a normal form game as represented in \ref{eqn:nf_game} where $I = {X_p, X_o}$ and $S$ is a set of all badminton shots

\begin{equation}
    G = <I, (s_i)_{i \epsilon I}, (u_i)_{i \epsilon I}>  \forall s_i, s_{-i} \epsilon S
    \label{eqn:nf_game}
\end{equation}

\begin{equation}
    u_{i}\left(s_{i} ,s_{-i}\right) =  P\left( s_{i}|s_{-i} \right)* P_{success}\left( s_{i}|s_{-i} \right)     
    \label{eqn:max_val}
\end{equation}

where $P( s_{i}|s_{-i} )$ is the probability of playing a shot $s_{i}$ for a given shot $s_{-i}$ of the opponent $X_o$, $P_{success}( s_{i}|s_{-i} )$ is the success rate of a shot $s_{i}$ for $s_{-i}$. These probabilities are calculated taking into account the data of the previous matches between the same two players. We consider the number of times a particular shot has been played by $X_p$ to calculate the probability of playing that shot including the instances where the shot yielded a point, resulted in the loss of a point or continuation of the rally. Within these instances, we consider the number of times that particular shot has yielded a point for $X_p$ while calculating the probability of success for that shot. This is done specifically for every shot played by $X_o$. For calculating the best response for a particular shot of $X_o$, we find the shot $s_{i}$ for $X_p$ which yields the maximum value of the utility according to the equation \ref{eqn:max_val}.

\begin{algorithm}
    \SetAlgoLined
    \KwResult{Shot recommendation $s_i$}
    \While{Point not gained}{
        1. Select $s_i$ for $X_p$ for given $s_{-i}$ from $X_o$ \\
        2. Calculate utility $u_i (s_i | s_{-i}), \forall s_i \epsilon S$ according to equation \ref{eqn:max_val} \\
        3. Pick $s_i$ with maximum utility $u_{max}$ \\
        4. Play $s_{i_{u_{max}}}$\\usepackage{}
    }
     \caption{Algorithm for recommendation system}
     \label{alg:rec_sys}
\end{algorithm}

This recommendation can be a return to a type of service or any other shot during the rally. It will help $X_p$ be prepared to face $X_o$ with confidence and certainly rule out few shots which have resulted in an immediate point loss in $X_p$'s history. Now, we can extend this tool further where we consider $X_o$ as our primary player and make all the computations to find his best possible shots for all the shots of $X_p$ based on the same data set. This will return a set of most probable shots of a particular for each shot of $X_p$. Now, this will help predict the return shot of $X_o$ for a particular shot $s_{i}$ of $X_p$. Accurate prediction of the type of serve or the type of return of $X_o$ for a particular shot $s_{i}$ of $X_p$ can prove to be very important in winning a point in crucial situations like dues or the first point of the set. This approach is the base for our other two methods and proves to be a vital one in real-world scenarios. We can observe many cases both from recommendation and the datasets, where a player rightly predicts a return of $X_o$ and surprises with a trick shot and wins a point.

\subsection{Simulator}
The simulator is an extension to the recommendation tool but modelled as an \textit{extensive form game} instead of a normal form game. The most important value addition to this method is that the history of shots between the players $X_p, X_o$ is taken into consideration. In badminton, it may not be always the case that the last shot results in winning or losing a point; the earlier shots played during the rally can also be responsible for a particular outcome. We observed from the collected data that this dependency on the history needs to be considered a maximum for two earlier shots for best results. We have introduced a reward system for incorporating history.
We consider four types of rewards as follows:
\begin{enumerate}
    \item A high positive reward $R_{hp}$ when a shot of $X_p$ results in a direct point. For instance, a smash resulting in a direct point; $R_{hp}$ = +5
    \item A medium positive reward $R_{mp}$ when a shot of $X_p$ induces a poor return from $X_o$ and thereby yielding a point. For instance, a good lift to the back making $X_o$ make a poor clearance helping $X_p$ kill immediately and gain a point; $R_{mp}$ = +2
    \item A medium negative reward $R_{mn}$ for a poor shot of $X_p$ which $X_o$ takes advantage of making $X_p$ lose a point; $R_{mn}$ = -2
    \item A low negative reward $R_{ln}$ when a shot results in a direct point loss; $R_{ln}$ = -5
\end{enumerate}

The rewards are hence considered with history up to two shots which will help the algorithm suggest the best possible outcome for $X_p$. The total reward for a shot of player $X_p$ given an opponent's shot is stated as follows:
\begin{equation}
\begin{split}
R_{T} (s_{i}, s_{-i}) = (R_{hp} (s_{i}, s_{-i})) + (R_{mp}(s_{i}, s_{-i}, s_{i}:t-1)) \\
+ (R_{mn}(s_{i}, s_{-i}, s_{i}:t-1)) + (R_{ln}(s_{i}, s_{-i}))
\end{split}
\end{equation}

The purpose of the simulator is to help $X_p$ by predicting the result of a rally up to a predefined number of steps through the match. It will help $X_p$ emulate the sequence of rallies in different ways to practice with another person before the match. Though the game of badminton is a perfect information zero sum game, the simulator is modelled as a \textit{a perfect-information infinite non-zero sum extensive form game} as the end utilities according to equation \ref{eqn:max_val} won't be the same for both the players. It is not possible to solve an infinite game. Hence we make a few modifications to the above-defined game into a finite extensive form game to be able to solve it up to a predefined number of steps. We call it \textbf{\textit{the approximated sequential representation}} of the original game. Here, we restrict the number of sequences to a predefined number depending on the type of sport we are applying to. In badminton, it is enough for a player to think about his next two moves with respect to one move of the opponent in between as he could rectify his mistakes within that else it would result in a point gain or loss within that. The scenario is illustrated in the figure \ref{Illustration}

\begin{figure*}
    \centering
    \includegraphics[width=7in]{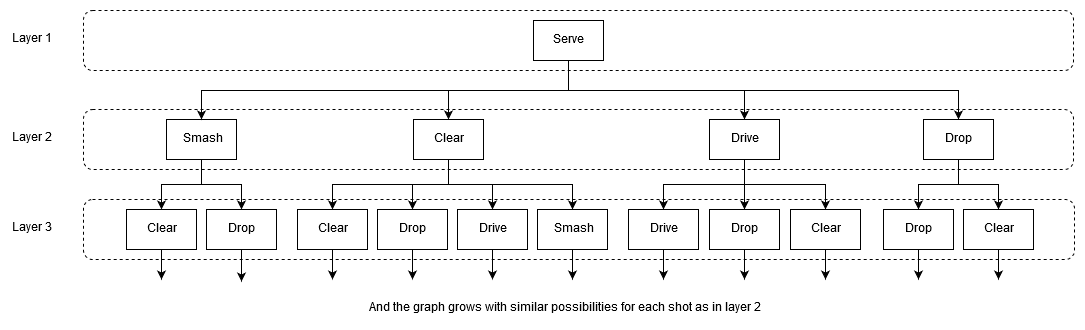}
    \caption{Extensive form representation of the game}
    \label{Illustration}
\end{figure*}

We introduce a reward system which is inspired from reinforcement learning \cite{article} \cite{KLMSurvey} that helps us calculate the favorable outcome for the player while taking into consideration the opponent’s moves. The model of the game can be represented as in equation \ref{eqn:sim_eqn} where $I = {X_p, X_o}$ is the set of agents (players), $S$ is the set of possible badminton shots, $H$ is the set of choice nodes, $Z$ is the set of terminal nodes, $\alpha$ agent function, $\beta$ action function and $\rho$ successor function respectively. This model is treated as a tree $T(n)$ where n is the number of nodes and $T$ can be expanded to $n$ levels denoting history of past events (3 in our case) for simulation. The game for $n$ levels is solved using \textit{backward induction} with data containing the updated reward values of shots depending on $(\alpha, \beta, \rho)$.

\begin{equation}
    G = <I, S, H, Z, \alpha, \beta_{i}, \rho, u_i>
    \label{eqn:sim_eqn}
\end{equation}

\begin{equation}
    u_{i}(s_{i} ,s_{-i} ) =  P( s_{i}|s_{-i} )* P_{success}( s_{i}|s_{-i} )* R_{T} ( s_{i},s_{-i} )
    \label{eq : er3}
\end{equation}

where $u_{i}$ is the utility of agent $i$, $P( s_{i}|s_{-i} )$ is the probability of $X_p$ playing a shot $s_{i}$ for a given shot $s_{-i}$ of $X_o$, $P_{success}( s_{i}|s_{-i} )$ is the success rate of a shot $s_{i}$ of $X_p$ for a given shot $s_{-i}$ of $X_o$, $R_{T}( s_{i},s_{-i})$  is the reward for playing $s_{i}$ for $s_{-i}$. By approximating the infinite extensive game into a finite extensive form game, based on the rewards and utilities of players, we can predict the progress in the game after each shot. The simulator then recommends the best favourable way the game could progress with the shots $X_p$ has to play along with the ones $X_o$ is expected to play. Though this model has the capability of producing very good results, it is often very hard to model the cognitive process of humans. In sports, humans tend to think differently, a move by a badminton player will involve a lot of factors like fatigue, ability, confidence, ability, condition in game, precision of opponents shot and even gut feeling. The algorithm for the simulation system is presented in algorithm\ref{alg:sim_sys}

\begin{algorithm}
    \SetAlgoLined
    \KwResult{Shot recommendation $s_i$}
    \While{Maximum number of predictions not reached}{
        1. Select $s_i$ for $X_p$ for given $s_{-i}$ from $X_o$ \\
        2. Expand simulation tree for next $n$ moves incorporating all possible shot combinations\\
        3. Calculate accumulated utilities for $X_p , X_o$ using equation \ref{eq : er3}\\
        4. Use backward induction to determine favourable shots and discard rest of the nodes\\
        5. Solve and reduce traversal path towards one optimal shot with maximum utility $s_{i_{u_{max}}}$\\
        6. Update values for all the shot types based on real time data\\
        7. Get $S_{-i}$ of $X_o$ for $s_{i_{u_{max}}}$ of $X_p$\\
    }
     \caption{Algorithm for simulator system}
     \label{alg:sim_sys}
\end{algorithm}

\section{Results}
In this section, we discuss the results of our models. As there is no established metric to verify the results of our proposed models, an expert opinion or domain knowledge is the only way to check the correctness and accuracy of results. As the model operates on the history of two players from the data, the results are only relevant to those players and will be different for others. Also, it is possible to create a generic model to focus on one player completely, given the data of the player with different opponents. The accuracy of the model depends on the volume and consistency of the available data.

\subsection{Results of the Recommendation System}
The recommendation system suggests the best response for an $X_o$’s shot without considering the history of shots in the rally and the condition of the player in the game. The results i.e. the recommendations of the model for the player $X_p$ are shown in Table 1 and that for the opponent $X_o$ are shown in Table 2. It can be seen from the data that the model has successfully identified the best shots for given shots and that it has discarded the shots that are not playable for the opponent’s shot successfully. The number of suggestions can be increased and we have fixed it to 2 so that $X_p$ always has an alternative. Also, from the suggestions above, it can be seen that the shot $forehand drop$ has been repeated the most. This coincides with the fact that both the players are successful and capable in playing forehand drop shot without any error which was evident from the matches. Using this model, a player can be mentally prepared on what shot to play to a given shot of the opponent that can either lead to a point or keep $X_p$ alive in the rally to avoid losing a point. As this is modelled directly from the capability and behavior of the players, this information will be of value to the players before the match.

\begin{table}
    \centering
    \resizebox{0.48\textwidth}{!}{
    \begin{tabular}{|c|c|c|} 
    \hline
    \textbf{Opponent's shot} & \textbf{suggestion 1} & \textbf{suggestion 2} \\
    \hline
    backhand\_short\_serve & forehand\_drop & backhand\_lift \\
    \hline
    backhand\_long\_serve	& forehand\_drop &	backhand\_drop \\
    \hline
    forehand\_drop & forehand\_drop & backhand\_lift \\
    \hline
    backhand\_drop & backhand\_drop & forehand\_lift \\
    \hline
    forehand\_kill & forehand\_lift & backhand\_lift \\
    \hline
    backhand\_kill & forehand\_long\_clear & forehand\_drop \\
    \hline
    jump\_crosscourt\_smash & block & forehand\_lift \\
    \hline
    normal\_smash & forehand\_drop & block \\
    \hline
    jump\_smash & block & backhand\_lift \\
    \hline
    forehand\_long\_clear & forehand\_drop & backhand\_lift \\
    \hline
    backhand\_short\_clear & jump\_crosscourt\_smash & forehand\_long\_clear \\
    \hline
    backhand\_long\_clear & forehand\_lift & normal\_crosscourt\_smash \\
    \hline
    crosscourt\_clear & jump\_smash & forehand\_lift \\
    \hline
    forehand\_drive & forehand\_drive & backhand\_lift \\
    \hline
    backhand\_drive & forehand\_drop & forehand\_lift \\
    \hline
    forehand\_lift & forehand\_drop & jump\_smash \\
    \hline
    backhand\_lift & forehand\_drop & normal\_smash \\
    \hline
\end{tabular}
}
\caption{Recommendations for $X_p$}
\label{table:1}
\end{table}

 \begin{table}
\centering
\resizebox{0.48\textwidth}{!}{
\begin{tabular}{|c|c|c|} 
    \hline
    Opponent's shot & suggestion 1 & suggestion 2 \\ 
    \hline
    backhand\_short\_serve & forehand\_drop & forehand\_lift \\
    \hline
    backhand\_long\_serve & normal\_smash & forehand\_drive \\
    \hline
    forehand\_drop & forehand\_drop & backhand\_lift \\
    \hline
    backhand\_drop & forehand\_drop & backhand\_drop \\
    \hline
    forehand\_kill & backhand\_lift & forehand\_drop \\
    \hline
    backhand\_kill & backhand\_lift & forehand\_lift \\
    \hline
    normal\_crosscourt\_smash & backhand\_lift & block \\
    \hline
    jump\_crosscourt\_smash & forehand\_drop & forehand\_lift \\
    \hline
    normal\_smash & block & forehand\_drop \\
    \hline
    jump\_smash & backhand\_lift & block \\
    \hline
    body\_smash & block & forehand\_lift' \\
    \hline
    forehand\_short\_clear & forehand\_drop & backhand\_drop \\
    \hline
    forehand\_long\_clear & normal\_smash & forehand\_long\_clear \\
    \hline
    backhand\_short\_clear & normal\_crosscourt\_smash & forehand\_drive \\
    \hline
    backhand\_long\_clear & jump\_smash & backhand\_drop \\
    \hline
    crosscourt\_clear & forehand\_drop & normal\_smash \\
    \hline
    forehand\_drive & forehand\_drive & forehand\_long\_clear \\
    \hline
    backhand\_drive & forehand\_drive & backhand\_drive \\
    \hline
    forehand\_lift & forehand\_drop & forehand\_lift \\
    \hline
    backhand\_lift & forehand\_drop & forehand\_drive \\
    \hline
\end{tabular}
}
\caption{Recommendations for $X_o$}
\label{table:2}
\end{table}

\begin{figure}
    \centering
    \includegraphics[width=0.47\textwidth]{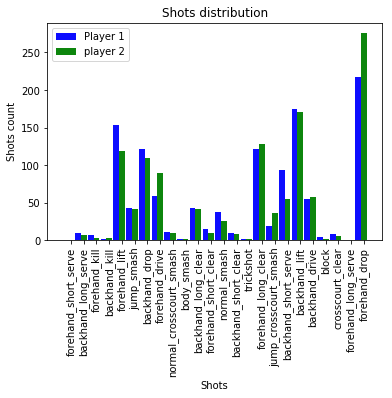}
    \caption{Frequency of shots for Player $P_{1}$ and player $P_{2}$ }
    \label{P12_shots}
\end{figure}

\subsection{Results of the simulator}
The purpose of the simulator is to take the history of the ongoing match into consideration and to model the game strategy for $X_p$ based on the position in the match, to identify the feasible sequence of shots that has lead to point gains and to avoid the poor shots. The simulator can be fed with seed shots i.e. few inputs in the start on how the game should proceed and it predicts the next few shots. Since there is no information about a point gain or loss to the simulator, it will infinitely predict the sequence of the shots in the game. It is accurate when compared to following only the best strategy since in reality, the player should think 2-3 steps ahead in badminton to realize the after-effect of playing a shot as there is always the possibility of a poor shot leading to a point loss during the consecutive shots played. Hence, at any point of time, the simulator builds a tree for the next three shots (2 for the player and 1 for the opponent), does backward induction on the utility values according to equation \ref{eq : er3}, arrives at the most favorable shot for $X_p$, discards the rest of the tree and the process is repeated. To check the working of the simulator, we seeded it with a few shots from a match's data which was not used for modelling. The actual sequence of shots from the match figure \ref{Illustration}  and the sequence of predicted shots from the simulator is given in Table \ref{table:3}.
 
\begin{figure}[h]
    \centering
    \includegraphics[width=0.48\textwidth]{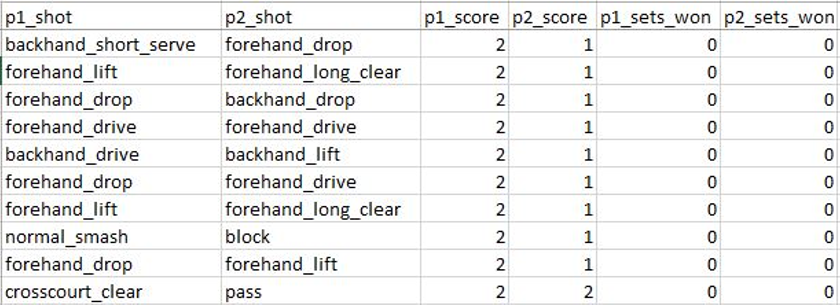}
    \caption{Data from an actual match between Lin Dan and Lee Chong Wei}
    \label{Illustration}
\end{figure}

\begin{table}[h]
    \centering
    \resizebox{0.48\textwidth}{!}{
    \begin{tabular}{|p{1cm}|c|c|}
    \hline
     Shot number & $X_p$'s (Lin Dan) shot & $X_o$'s (Lee Chong Wei) shot \\ 
     \hline
    1 & backhand\_short\_serve & forehand\_drop \\
    \hline
    2 & forehand\_lift & forehand\_short\_clear \\
    \hline
    3 & forehand\_drop & backhand\_drop \\
    \hline
    4 & forehand\_drive & forehand\_drop \\
    \hline
    5 & forehand\_drop & forehand\_drive \\
    \hline
    6 & backhand\_drop & forehand\_drop \\
    \hline
    7 & forehand\_drop & forehand\_long\_clear \\
    \hline
    8 & jump\_smash & forehand\_drop \\
    \hline
    9 & forehand\_drop & backhand\_short\_clear \\
    \hline
    10 & jump\_crosscourt\_smash & forehand\_drop \\
    \hline
\end{tabular}
}
\caption{Output from the Simulator}
\label{table:3}
\end{table}

It can be seen from the tables that the results of simulator are closer to the actual match. The results mostly differ only in the sub-category of shots and the actual type of shots are the same. This tool can help the players understand how the match will proceed after a shot is played which is crucial to analyze the repercussions of playing a shot. The simulator model is likely to work better when the data used for modelling is larger. Fig. \ref{P12_shots} shows the distribution of shots for both the players in the data collected. There is an unequal distribution of shots which affects the accuracy of predictions of the models. We assume that data from at least 20 full matches is required to precisely model the behavior as in our case with 3 matches, the frequency of most of the shots is very low. Almost equal distribution of shots is required to ensure the reliability of the results from the model.

\section{Conclusion}
In this paper, we have successfully developed two novel approaches for the development of an assistance tool for the game of badminton based on the concepts of Game Theory. Our recommendation tool takes in match data for the player under consideration against a particular opponent and gives out the best possible set of strategies (shots) which the player can use. The simulator model is a generalized and robust extension of this recommendation tool which considers the history of shots played in the ongoing match along with match history to suggest the favorable strategy for the players. The results, analysis and comparison with the actual match data shows the effectiveness of the system and that it is well-rounded. Our current work is restricted by the availability of data and the using the manually annotated data from 3 matches for our experiments was to show the feasibility of our approaches. In the presence of a considerable amount of annotated data, our future works are to test our approaches for other two player sports, remodelling the approaches for team sports, building a complete pipeline of dedicated software application or program that can dynamically function by adapting to the real-time change during a course of a tournament or a match using computer vision to capture visual information and reinforcement learning approaches like Markov Decision Process (MDP) to mathematically model decisions for a system of advanced assistance.

\bibliographystyle{ieeetr}
\bibliography{references}

\end{document}